\documentclass[aps,floatfix,amsmath,ams,prl
symb, twocolumn, showpacs]{revtex4}
\usepackage{graphicx}
\begin{document}

\title{Multifractal eigenstates of quantum chaos and the Thue-Morse sequence}
\author{N. Meenakshisundaram, Arul Lakshminarayan}
\affiliation{Department of Physics\\ Indian Institute of Technology Madras\\
Chennai, 600036, India.}


\begin{abstract}
We analyze certain
eigenstates of the quantum baker's map and demonstrate, using the Walsh-Hadamard 
transform, the emergence of the ubiquitous Thue-Morse sequence, a simple sequence
 that is at the
border between quasi-periodicity and chaos, and hence is a good
paradigm for quantum chaotic states. We show a family of
states that are also simply related to Thue-Morse sequence, and
 are strongly scarred by short periodic orbits and their
homoclinic excursions. We give approximate expressions for these states
and provide evidence that these and other generic states are multifractal.
\pacs{05.45.Mt, 05.45.Df}
\end{abstract}

\maketitle
\newcommand{\newc}{\newcommand}
\newc{\beq}{\begin{equation}}
\newc{\eeq}{\end{equation}}
\newc{\kt}{\rangle}
\newc{\br}{\langle}
\newc{\beqa}{\begin{eqnarray}}
\newc{\eeqa}{\end{eqnarray}}
\newc{\pr}{\prime}
\newc{\longra}{\longrightarrow}
\newc{\ot}{\otimes}
\newc{\rarrow}{\rightarrow}
\newc{\h}{\hat}
\newc{\bom}{\boldmath}
\newc{\btd}{\bigtriangledown}
\newc{\al}{\alpha}
\newc{\be}{\beta}
\newc{\ld}{\lambda}
\newc{\sg}{\sigma}
\newc{\p}{\psi}
\newc{\eps}{\epsilon}
\newc{\om}{\omega}
\newc{\mb}{\mbox}
\newc{\tm}{\times}
\newc{\hu}{\hat{u}}
\newc{\hv}{\hat{v}}
\newc{\hk}{\hat{K}}
\newc{\ra}{\rightarrow}
\newc{\non}{\nonumber}
\newc{\ul}{\underline}
\newc{\hs}{\hspace}
\newc{\longla}{\longleftarrow}
\newc{\ts}{\textstyle}
\newc{\f}{\frac}
\newc{\df}{\dfrac}
\newc{\ovl}{\overline}
\newc{\bc}{\begin{center}}
\newc{\ec}{\end{center}}
\newc{\dg}{\dagger}
\newc{\prh}{\mbox{PR}_H}
\newc{\prq}{\mbox{PR}_q}

 The emergence of nonlinearity induced chaos from the 
linear Schr\"{o}dinger equation in the classical
limit is a subtle and interesting phenomenon \cite{BerryQC,Heller,Haake}.
The eigenstates of quantized chaotic systems are of current interest. They range from those
that bear striking resemblance to classical periodic orbits, and hence
called ``scarred states'' \cite{Heller}, to those that are fruitfully
 described by the statistical properties of random matrices \cite{Haake}.
  Simple models, such as the logistic map, reveal the essence of classical chaos \cite{LL}.
Comparable quantum models, such as the quantum cat map \cite{Cat1} and
 the quantum baker's map \cite{BalVor}, while providing valuable insights,
  are less tractable.  

The classical baker's map \cite{LL}, $T$, is the area preserving transformation
of the unit square $[0,1)\times [0,1)$ onto itself, which takes a
phase space point $(q,p)$ to $(q',p')$ where $(q'=2q,\, p'=p/2)$ if
$0\le q<1/2$ and $(q'=2q-1,\,p'=(p+1)/2)$ if $1/2\le q<1$.  The
stretching along the horizontal $q$ direction by a factor of two is
compensated exactly by a compression in the vertical $p$
direction. The repeated action of $T$ on the square leaves the phase
space mixed, this is well known to be a fully chaotic system that in a
mathematically precise sense is as random as a coin toss. The
area-preserving property makes this map a model of chaotic two-degree
of freedom Hamiltonian systems, and the Lyapunov exponent is
$\log(2)$ per iteration.  

The baker's map was quantized by Balazs and Voros \cite{BalVor}, while 
Saraceno \cite{Saraceno} imposed anti-periodic boundary conditions. 
We consider exclusively the case $N=2^K$ for integer $K$.
 It has been known from early on that this case is special for the
  quantum baker's map \cite{BalVor}, but this has not so far been made use
   of to analyze eigenstates more closely. The quantum baker's map,
    in the position representation, that we use in this Letter is:
 \beq B=G_K^{-1}\left( \begin{array}{cc}G_{K-1} &0
\\0 & G_{K-1} \end{array}\right), \eeq where $(G_{K})_{mn}=\br
p_m|q_n\kt = \exp[-2 \pi i (m+1/2)(n+1/2)/N]/\sqrt{N}$.
The Hilbert space is finite dimensional, the dimensionality $N$ being the 
scaled inverse Planck constant $(N=1/h)$, where we have used that the phase-space
area is unity. The position and momentum states are denoted as $|q_n\kt$ and $|p_m\kt$,
where $m,n=0,\cdots,N-1$ and the transformation function between these bases is the
finite Fourier transform $G_K$ given above.
 The choice of anti-periodic boundary conditions fully preserves parity
  symmetry, here called $R$, which is such that $R|q_{n}\kt = |q_{N-n-1}\kt.$ 
  Time-reversal symmetry is also present and implies in the context of
  the quantum baker's map that an overall phase can be chosen such that the
momentum and position representations are complex conjugates: $G_K
\phi=\phi^{*}$, if $\phi$ is an eigenstate in the position basis. 
$B$ is an unitary matrix, whose repeated application is the quantum version
 of the full left-shift of classical chaos. There is a semiclassical trace formula,
 which, based on the unstable periodic orbits, approximates eigenvalues \cite{Almeida}.

\begin{figure}
\label{PR}
  \includegraphics[width=2.75in]{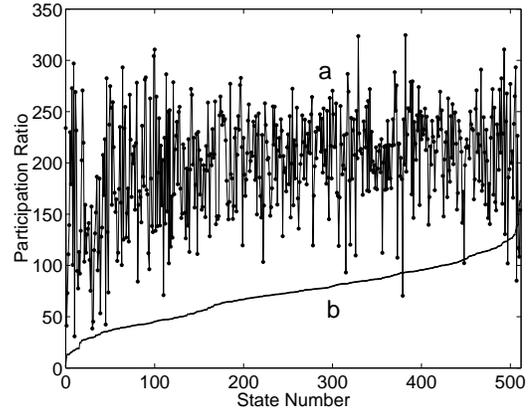}
  \caption{The participation ratio of the eigenstates of the quantum baker's map
 for $N=512$ in the (a) position basis, and (b) Walsh-Hadamard basis.
 The states are arranged in increasing order of the latter number.} 
\end{figure}  

   Our first step
in identifying special states is to Walsh-Hadamard (WH) transform \cite{Schroeder} 
eigenstates; if $\phi$ is an eigenstate we study $H_K\phi$, where
$H_K=\otimes^K H$, a $K$-fold tensor product of the Hadamard matrix $H=((1,1),(1,-1))/\sqrt{2}$.
  Very significant simplifications are seen in this basis for almost all
  states. In particular we analyze the number of principal components
  in the WH transform by means of the participation ratio $\prh$ which is
  $1/\sum_{i}|(H_K\phi)_i|^4$.  In Fig.~1 we compare this participation ratio with that calculated
  in the original position basis. For one remarkable class of states that
  is present for all $N$, powers of two, $\prh$ is the smallest and
  of the order of unity, for example when $N=1024$, $\prh=1.96$ for this state.
   This class of states seems to have been identified by Balazs and Voros \cite{BalVor}, as the
  eigenvalue is very close to $-i$, however, the WH transform reveals
  their simplicity and beauty. In Fig.~2 are examples of these states, and their WH transform,
   where we also see their near self-similarity. We call these set of states, $\phi_{tm}$ the ``Thue-Morse''
    states, as the principal peak in the Hadamard transform corresponds to the 
    overlap with the final column (or row) of $H_K$ which (apart from the factor $1/\sqrt{2^K}$)
     is the $K$-th generation of the Thue-Morse sequence. 

\begin{figure}
\label{TM}
\includegraphics[width=2.75in]{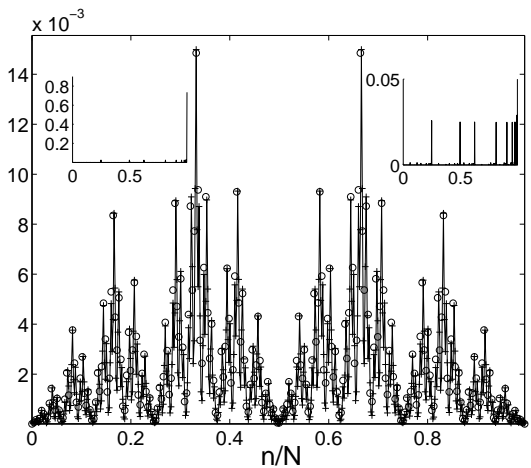}
\caption{ The intensity $|\br q_n|\phi_{tm}\kt|^2$ of the Thue-Morse state in the position
 representation. Shown are the states for $N=512$ (plus) and $N=256$ (circle) after 
 the latter state was appropriately scaled. Also shown is the estimate $\phi_{tm}^A$ for
  $N=512$ case (line). The left inset is the intensity of the WH tranform of the
  Thue-Morse state for $N=512$, while the right inset shows the same on a different scale. }
\end{figure}
  
The Thue-Morse sequence is an example of a binary ``automatic sequence'' \cite{Allpaper},
whose first few terms are $\{1,-1,-1,1,-1,1,1,-1,\cdots\}$.
The finite generations are constructed
as follows: start with $t_0=\{1\}$ and generate $t_1=\{1,\,-1\}$ by
appending $\overline{t_0}$ to $t_0$, where the overbar is
multiplication by $-1$, and we continue to iterate the rule $t_{k+1}=
\{t_{k}, \,\overline{t_{k}}\}$.  At stage $K$ we get $t_{K}$ a string
of length $2^K$, which we also treat as a column vector whose $n$-th
element we denote as $t_K(n)$. The concatenation rule above is then
equivalent to the generating rules: $t_K(2n)=t_{K-1}(n)$ and
$t_{K}(2n+1)=-t_{K-1}(n)$.  The Thue-Morse sequence considered as a
word with two alphabets is cube free, that is no block repeats thrice
consecutively. This sequence occurs in numerous contexts \cite{Allpaper},
and is marginal between a quasiperiodic sequence and a chaotic one. The deterministic
disorder of this sequence is relevant to models of quasicrystals \cite{Bovier} and 
mesoscopic disordered systems \cite{Janssen}, and as we show here quantum chaos. We
believe that this is the first time it has emerged naturally in a
quantum mechanical problem, rather than being assumed.

That $t_K$ is an approximate eigenvector of $B$ follows on using the concatenation rule to
get $(B\,t_K)_l\,=\,$
\[ \f{\sqrt{2}}{N} C_l\,\sum_{m=0}^{N/2-1} t_{K-1}(m) \sum_{n=0}^{N/2-1}e^{\f{2\pi i}{N}
 (n+1/2)(l-2m-1/2)},\]
where $C_l=(1-i (-1)^l)$. For large $N$ the $n$ sum sharply peaks at $l=2m$, or $2m+1$ depending on
 if $l$ is even or odd. Thus $(Bt_K)_{2m}\approx -i (2\sqrt{2}/\pi) t_{K-1}(m)$ and
 $(Bt_K)_{2m+1}\approx i (2\sqrt{2}/\pi) t_{K-1}(m)$, from which using the generating
 rules of the Thue-Morse sequence we get $(B\,t_K)_m \approx -i (2\sqrt{2}/\pi) t_K(m)$.
 The eigenvalue we get has a modulus $\approx 0.9$.
We can alternatively calculate exactly $G_K
t_K$ as a product if we use the fact that $t_K(n)$ is $(-1)^r$ where
$r$ is the number of times that $1$ occurs in the binary
representation of $n$. A short calculation gives $(G_K\, t_K)_m\,=\,$
 \[
 \sqrt{2^K}
 (-1)^{m+1}\,(i)^{K+1}\prod_{j=0}^{K-1}\sin[\pi
 2^{j-K}(m+1/2)],
 \]
a useful formula from which also we can see that $t_K$ is an approximate eigenvector of $B$.

For example when $N=2^9$, $|\br t_9|\phi_{tm}\kt |^2/N \approx 0.74$.
 Taking into account symmetries helps us do much better. It is easy to
see that $R\,t_K=(-1)^{K}t_K,$ and indeed the parity of the Thue-Morse state
flips with each power of two.  We note however that $G_K\, t_K \ne
t_K^*$ and therefore $t_K$ as such does not have the correct symmetry
properties. To take advantage of time-reversal symmetry we construct
$\phi_0=\gamma \,t_K + \gamma^{*}G_K^{-1} t_K$, which satisfies
$G_K\phi_0=\phi_0^*$. It is straightforward to see that  $G_K^{-1}t_K$
 is also an approximate eigenstate of $B$. We determine
 numerically the complex constant $\gamma$ such that $\phi_0$ best approximates the 
 state $\phi_{tm}$. For example if $N=512$, $\gamma$ is predominantly real
and just $\phi_0=C\,(t_9+G_{9}t_9)$ is such
 that $|\br \phi_0|\phi_{tm}\kt|^2 \approx 0.93,$ that is the state $\phi_{tm}$
 is determined to more
 than $93\%$ by the Thue-Morse sequence and its Fourier transform.
$C$ is a normalization factor and we have used that $G_{9}^{-1}t_9=G_{9}t_9$.
Thus the determinisitic structural disorder of the Thue-Morse sequence is seen in a 
quantized model of a classically deterministic and fully chaotic system.

 We note that we can 
 improve upon the simple ansatz $\phi_0$ above, by taking into account the second rung of 
 peaks in the Hadamard transform of $\phi_{tm}$, $K$ easily identifiable peaks ($N=2^K$), 
 as seen for example in the right inset of Fig.~2.
 These second rung of peaks also result from the Fourier transform of the Thue-Morse
 sequence, but not exclusively. We introduce the notation $t^{(r)}_k$ for an $2^r$-fold repetition of the Thue-Morse sequence
 of generation $k$, for example column number $2^{K-r}-1$ of $H_K$ is $t^{(r)}_{K-r}$.
   The improved ansatz for the Thue-Morse state is then
  $\phi_{tm}^A= \gamma_0\phi_0+\sum_{k=1}^{K} (\gamma_k\,
    +\, \gamma^*_k\,G_K^{-1})S^{k-1} \, t^{(2)}_{K-2},$ where $S$ is the shift operator, and 
we again determine the complex constants $\gamma_k$ numerically such that the ansatz is 
closest to the state. 
    
 The shift operator $S$, acts on the position basis as $S|q_n\kt = |q_{2n}\kt$ or
 $|q_{2n-N+1}\kt$ depending on if $n< N/2$ or otherwise. We notice that $S$ is ``almost" $B$, only
 there is no momentum cut-off, as $\br p_m|B|q_n\kt = \sqrt{2}\br p_m|q_{2n}\kt$ for $n$ and $m$ both $\le N/2-1$. Infact $S$ was proposed as the quantum baker's map
  by O. Penrose \cite{Penrose},
  however, apart from other issues, it does not respect time-reversal symmetry, 
  in the sense that $G^{-1}_KS^*G_K \ne S^{-1}.$  If the position
  state $|q_n\kt$ is denoted in terms of the binary expansion of $n=a_{K-1}a_{K-2}\cdots a_0$
   then $S|a_{K-1}a_{K-2}\cdots a_0\kt=|a_{K-2}a_{K-3}\cdots a_0 a_{K-1}\kt$, hence $S^K=1_N$.
    Such shift operators including momentum like bits were recently proposed as quantum baker's maps,
     however these also do not seem to respect time reversal symmetry \cite{SchackCaves}.
   We can find $\gamma_k$ such that $|\br \phi_{tm}|\phi_{tm}^A\kt|^2 \approx 0.998$, this is 
shown in Fig.~2. The emergence of the shift operator is unsurprising as it commutes with the Hadamard
transform $H_K$, and is close to the quantum baker's map $B$. The Thue-Morse sequence $t_K$ is an
exact eigenstate of $S$ and is hence a stand-alone state.

It is known that the support of the spectral measure of the infinite Thue-Morse
sequence is a multifractal \cite{Luck}, thus it comes as no surprise that $\phi_{tm}$ has multifractal
properties. We can study these either by comparing Thue-Morse states at different values
of $K$ or for a fixed $K$, by averaging over different sized bins. We perform a standard
 multifractal analysis \cite{Halsey} and
see typical $f(\alpha)$ singularity spectrum. One simple diagnostic is the scaling of the
inverse participation ratio (IPR). If $\phi_n$ denotes the position representation of the
state $\phi$, the IPR which is $\sum_{n=0}^{N-1}\,|\phi_n|^4$ scales as $N^{-D_2}$. When
 $D_2$, the correlation dimension, is such that $0<D_2<1$, the function is a multifractal,
 between localized states ($D_2=0$) and random states ($D_2=1$). This IPR and multifractal analysis is
 carried out in the physically relevant position basis. Either of the two procedures gives for the Thue-Morse state the dimension $D_2 \approx 0.86$,
  indicating its multifractal character. This scaling along with the full $f(\alpha)$ spectrum
  is shown in Fig.~3.
  
 \begin{figure}
 \label{D2}
 \includegraphics[width=2.75in]{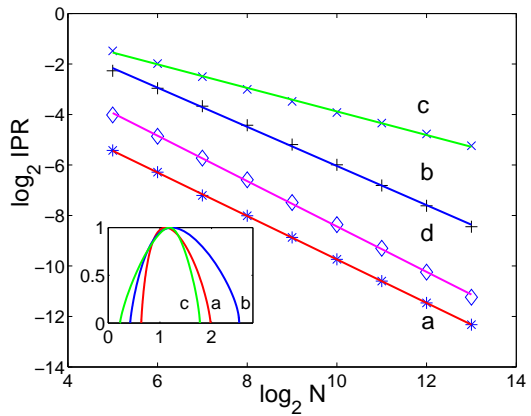}
 \caption{(Color Online) The scaling of the IPR in the position basis as a function of $N$.
  Shown as points are the numerical data,
   while the smooth curves are the best fit straight lines. The cases are:
  (a) the Thue-Morse state, (b) a period$-2$ scarred state, 
  (c) fixed-point scarred state after projecting out the uniform state,
  and (d) average over all the states. The corresponding $f(\alpha)$ spectrums 
  for the case $N=8192$ are shown in the inset.}
 \end{figure}

    The Thue-Morse state is by far not the only one influenced by the Thue-Morse sequence.   
We now construct a family of strongly scarred states consisting of $K$ members.
The $\prh$ is low for these states also, for example when $N=512$ $\prh \sim 15$.
We identify this as a ``family'' based on the similarity of their $\prh$ and the
 shared position of the $K$ principal peaks in their WH  
transform. For up to $N=2^{13}$ we have verified that this family exists and has $K$ 
prominent peaks, at $t^{(1)}_{K-1}$ and all its $K-1$ shifts, $S^k\, t^{(1)}_{K-1}.$
Amongst this family, members are strongly scarred by period$-2$, period$-4$ and related 
homoclinic orbits. Two examples of the members of this family along with their WH
 transforms are shown in Fig.~4. 
Taking time-reversal symmetry into account we may then write an approximate expression for
 this family of states based on $K$ vectors, similar to the one above for the Thue-Morse state.
For example for $N=512$ this procedure enables us to reproduce the state strongly scarred 
by the period$-2$ orbit up to more than $95\%$ and most of the family to similar accuracy.
The existence of such families may be indicative of systematics of periodic orbit scarring.

\begin{figure}
\includegraphics[width=2.75in]{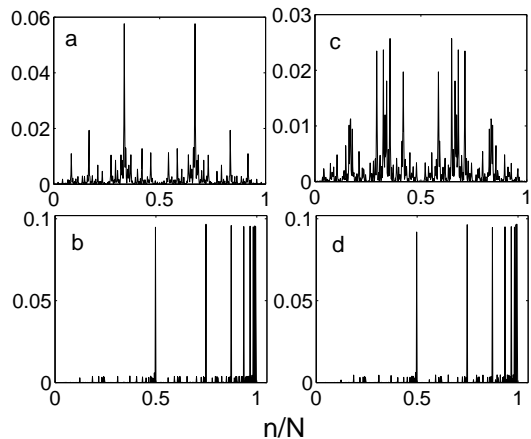}
\caption{ Two members of a family scarred by the period-2 orbit for $N=512$. 
Shown are the intensities (in position basis) of the states in (a) and (c), while
the corresponding WH transforms are shown in (b) and (d).}
\end{figure}

This strongly scarred state and its family members also have multifractal properties, and indeed
$D_2 \approx 0.8 $ for the period$-2$ scarred state (Fig.~3). We have identified other families,
 but we now turn to the state that is strongly scarred by the fixed point $(0,0)$ and whose
WH transform is strongly peaked, indeed corresponding to $t^{(K)}_0/\sqrt{2^K}$, that is simply
the uniform state. This also forms a series, for different $N$, whose eigenangles
 are close to zero. Since the Fourier transform of the uniform state given above tends
  to a constant (independent of $N$) at the origin (and at $1$), we expect
   the scarred state to be dominated by this ``Bragg peak''.
  In order to see any multifractal character even in this state we project $t^{(K)}_0$ out of
   its WH transform and analyze the resultant
  state (after renormalizing) in the position basis. This shows that the ``grass'' even in this
  strongly scarred state is a multifractal with a value of $D_2\approx 0.45$, as shown in Fig.~3,
  where the $f(\alpha)$ spectrum for the fixed point and period$-2$
  scarred states are also shown conforming to those expected of multifractals.
  For a gross measure of the multifractality of the entire spectrum, we
averaged the IPR (in the position basis) of all the states and found a
scaling with $D_2 \approx 0.9$, that is shown in Fig.~3.

    Classical structures in quantum states are clearly seen in their phase-space 
  or coherent state representations \cite{Saraceno} which is shown in Fig.~5 for
   the Thue-Morse states we have discussed above. Note the self-similarity of the four
   states shown here, the two different types of intricate structures arise
  from the alternating parity of these states. While dominated by the period-2 orbit and its
  associated structures, we leave the details for a later work.
   
\begin{figure}
\label{husimi}
\includegraphics[width=3.25in]{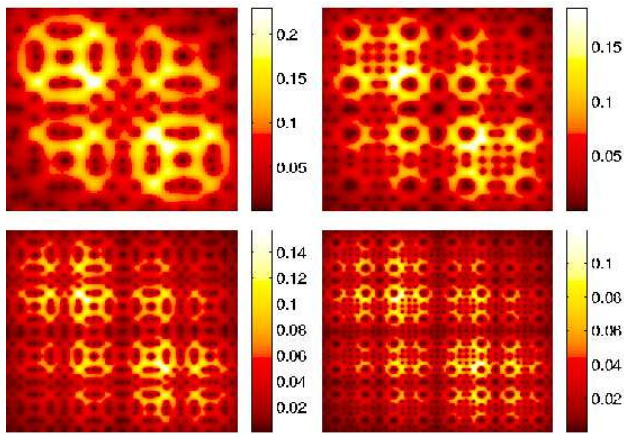}
\caption{(Color Online) The density plots of $|\br qp|\psi\kt|$ on the unit
  phase-space square for the Thue-Morse state in the cases $N=128,\,
256,\, 512,$ and $1024$ (left to right, top to bottom).}
\end{figure}

 We have
merely skimmed the surface of the states in this Letter, highlighting
the novel features found. The classical baker's map 
straightens out the intricate homoclinic tangle of unstable and
stable manifolds to lie along the $q$ and $p$ directions, which are
also the basis in which we are viewing the quantum states. The
multifractality and the simplicity of the states may originate in
this, while in other models the complexity of the tangle may give rise
to a ``mixing'' that makes them look random. That the dimensionality
is a power of two is also a critical requirement for the observations
in this Letter. Qubit implementation of the quantum baker's map, such as
the experimental work with NMR bits \cite{NMR} are naturally of this kind.
However self-similarity has been noticed indirectly
earlier in the quantum baker's map in the correlations between the
eigenstates and level-velocities \cite{Arul}, even for $N$ that are
not powers of two.  

 \acknowledgments{
N.M. was supported by financial assistance from the Council for Scientific and
 Industrial Research, India. We thank H. R. Naren for involvement in the early
  stages of this work.}

\end{document}